# Thermodynamics analysis on BaF$_2$ intermediate phase in solution-derived YBCO superconducting film deposition

Feng Feng[1], Hongyuan Lu[1], Wei Wu*[2], Xiangsong Zhang[1], Linli Wang[1]

[1] Division of Advanced Manufacturing, Graduate School at Shenzhen, Tsinghua University, Shenzhen 518055, China
[2] Department of Electrical Engineering, Shanghai Jiao Tong University, Shanghai 200240, China

**Abstract:** In the YBa$_2$Cu$_3$O$_{7-\delta}$ (YBCO) high temperature superconducting thin film fabrication via the chemical solution deposition method, BaF$_2$ is an important intermediate phase during heat treatment. In this paper, BaF$_2$ thermodynamics stability was analyzed through calculating the standard Gibbs free energy change ($\Delta G_T$) of the reactions related to other intermediate phases within the temperature range of 700-1000 K. Two thermodynamics methods, the Gibbs free energy function method and standard formation molar Gibbs free energy method, were utilized to obtain the $\Delta G_T$ values. The formation priority of BaF$_2$ relative to other intermediate phases were verified at higher temperatures, while the possibility of BaCO$_3$ formation was found at 700 K.

REBa$_2$Cu$_3$O$_{7-\delta}$ (REBCO, RE represents Rare Earth elements including Y, Gd, Sm, etc.) high temperature superconducting (HTS) materials are of important application potentials in many fields [1, 2]. The precursor solution containing trifluoroacetate (TFA) is widely used to fabricate of YBCO HTS thin films, which is named as the metal organic deposition (MOD) route [3, 4]. In the MOD-YBCO process, the precursor solution is coated on the substrate and heat treated, and the heat treatment usually consists of pyrolysis, crystallization and oxygenation [5, 6]. After the pyrolysis step, the organic precursor salts are decomposed into intermediate phases, which could evolve along with the temperature raising and finally react into YBCO phase during the crystallization step. Among the intermediate phases, BaCO$_3$ is detrimental since it is stable at the crystallization temperature, degrading the superconducting performance of the final YBCO film. BaF$_2$ is generally regarded as an important intermediate phase to suppress BaCO$_3$ formation [6, 7].

In the study of Clem [8], the thermodynamics stabilities of the BaCO$_3$/BaF$_2$/BaTiO$_3$ perovskite system was calculated using Thermocalc$^{TM}$, and BaF$_2$ was predicted to be more stable than BaCO$_3$ at temperatures ranging in 0-1000°C. In this study, the BaF$_2$ stability was calculated by considering possible conversion reactions, which involved other intermediate phases, such as YF$_3$ and CuO. In order to obtain more understandings of the BaF$_2$–related reactions, two methods were used for the calculation of standard Gibbs free energy change ($\Delta G_T$) at the temperatures ($T$) of 700, 800, 900 and 1000 K, which was within the $T$ range of intermediate phase evolution.

## 1 Analysis Methods

### 1.1 Intermediate phase conversion reactions

In order to investigate the thermodynamics stability of BaF$_2$, the Ba-related and F-related phases would be considered as possible intermediate phases, including BaO, BaCO$_3$, Ba(OH)$_2$, YF$_3$ and CuF$_2$. It could be noted that there are also some other related intermediate phases such as Y$_2$Cu$_2$O$_5$, Ba$_{1-x}$Y$_x$F$_{2+x}$, Ba-O-F [5, 6]. However, their thermodynamics data were not available in handbooks, thus these phases would not be included in this study. The possible conversion reactions involving the above phases could be summarize as Table 1.

Table 1 Possible conversion reactions of BaF$_2$.

| Reaction label | Reaction equation |
|---|---|
| (a) | BaF$_2$+CuO→BaO+CuF$_2$ |
| (b) | 3BaF$_2$+Y$_2$O$_3$→3BaO+2YF$_3$ |
| (c) | 3BaF$_2$+Y$_2$O$_3$+3CO$_2$(g)→3BaCO$_3$+2YF$_3$ |
| (d) | BaF$_2$+CuO+CO$_2$(g)→BaCO$_3$+CuF$_2$ |
| (e) | 3BaF$_2$+Y$_2$O$_3$+3H$_2$O(g)→3Ba(OH)$_2$+2YF$_3$ |
| (f) | BaF$_2$+CuO+H$_2$O(g)→Ba(OH)$_2$+CuF$_2$ |

A high positive $\Delta G_T$ value of a certain reaction would indicate that the reaction could not occur in the standard condition. Thus the $\Delta G_T$ calculation of the listed reactions could be used to interpret the possibility of BaF$_2$ conversion, and investigate the formation priority of BaF$_2$ during the intermediate phase evolution. In this study, two methods of $\Delta G_T$ calculation would be used, which were the Gibbs free energy function method and standard formation molar Gibbs free energy method, respectively. We would combine and compare the results calculated using the two methods below.

### 1.2 Gibbs free energy function method

In the Gibbs free energy function method through a series of transformation introduced by Ye [9], the $\Delta G_T$ calculation was transformed to the Gibbs free energy function ($\phi_{i,T}$) of phase $i$, whose stoichiometric coefficient in the reaction equation was $n_i$, and the calculation of standard molar formation enthalpy ($\Delta H_{i,f,298}$) values, as shown in equations (1-3).

$$\Delta H_{298} = \sum (n_i \Delta H_{i,f,298})_{products} - \sum (n_i \Delta H_{i,f,298})_{reactants} \quad (1)$$

$$\Delta \phi_T = \sum (n_i \phi_{i,T})_{products} - \sum (n_i \phi_{i,T})_{reactants} \quad (2)$$

$$\Delta G_T = \Delta H_{298} - T \Delta \phi_T \quad (3)$$

Table 2 Gibbs free energy function $\phi_{i,T}$ and standard molar formation enthalpy $\Delta H_{i,f,298}$ in reference [9], the unit is J mol$^{-1}$

| Phases | $\Delta H_{f,298}$ | $\phi_{700}$ | $\phi_{800}$ | $\phi_{900}$ | $\phi_{1000}$ |
|---|---|---|---|---|---|
| BaF$_2$ | -1207084 | 118.279 | 124.868 | 131.210 | 137.268 |
| CuO | -155854 | 55.690 | 59.701 | 63.604 | 67.374 |
| BaO | -553543 | 84.164 | 88.330 | 92.343 | 96.178 |
| CuF$_2$ | -548941 | 89.147 | 95.266 | 101.181 | 106.866 |
| Y$_2$O$_3$ | -1905394 | 130.667 | 140.105 | 149.182 | 157.845 |
| YF$_3$ | -1718369 | 137.227 | 145.319 | 153.068 | 160.439 |
| CO$_2$(g) | -393505 | 225.440 | 229.058 | 232.568 | 235.946 |
| BaCO$_3$ | -1216289 | 139.866 | 148.475 | 156.898 | 165.077 |
| H$_2$O(g) | -241814 | 198.413 | 201.285 | 204.057 | 206.716 |
| Ba(OH)$_2$ | -943492 | 127.457 | 135.237 | 143.017 | 150.797 |

The $\phi_{i,T}$ and $\Delta H_{i,f,298}$ data of the phases involved in reactions (a) - (f) of Table 1 could be found in the reference [9], and were listed in Table 2. It should be noted that the melting point of Ba(OH)$_2$ is 681 K, while its $\phi_T$ data could only be found within the $T$ range of 298~681 K, where Ba(OH)$_2$ is of solid state. Therefore, the values listed in Table 2 was estimated by linear extrapolation using the known data, which would result in a certain deviation out of the real values since Ba(OH)$_2$ should be of liquid state.

### 1.3 Standard formation molar Gibbs free energy method

The $\Delta G_T$ calculation method using standard formation molar Gibbs free energy ($\Delta G_{f,T}$) are introduces in many thermodynamics textbooks, with the calculation procedure shown in equation (4) [10]. $\Delta G_{f,T}$ values of the related phases were cited out of reference [11], as shown in Table 3.

$$\Delta G_T = \sum (n_i \Delta G_{f,T})_{products} - \sum (n_i \Delta G_{f,T})_{reactants} \quad (4)$$

Table 3 Standard formation molar Gibbs free energy $\Delta G_{f,T}$ in reference [11], the unit is kJ mol$^{-1}$

| Phases | $\Delta G_{f,700}$ | $\Delta G_{f,800}$ | $\Delta G_{f,900}$ | $\Delta G_{f,1000}$ |
|---|---|---|---|---|
| BaCO$_3$ | -1033 | -1007 | -981 | -956 |
| BaF$_2$ | -1092 | -1076 | -1060 | -1044 |
| BaO | -487 | -478 | -468 | -458 |
| Ba(OH)$_2$ | -746 | -722 | -697 | -673 |
| Y$_2$O$_3$ | -1699 | -1671 | -1643 | -1615 |
| YF$_3$ | -1540 | -1516 | -1492 | -1468 |
| CuO | -92 | -83 | -75 | -66 |
| CuF$_2$ | -431 | -416 | -402 | -389 |
| CO$_2$(g) | -395 | -396 | -396 | -396 |
| H$_2$O(g) | -209 | -204 | -198 | -193 |

## 2 Results and Discussion

### 2.1 Comparison of $T$ dependence Gibbs free energy change calculated using two methods

The Gibbs free energy change ($\Delta G_T$) of the reactions in Table 1 were calculated using the two methods mentioned above, and the obtained results were listed in Tables 4 and 5, respectively.

Table 4 Gibbs free energy change $\Delta G_T$ of the reactions in Table 1 calculated using Gibbs free energy function method, the unit is kJ mol$^{-1}$

| label | $\Delta G_{700}$ | $\Delta G_{800}$ | $\Delta G_{900}$ | $\Delta G_{1000}$ |
|---|---|---|---|---|
| (a) | 261 | 261 | 262 | 262 |
| (b) | 400 | 397 | 393 | 390 |
| (c) | -51 | -6 | 39 | 83 |
| (d) | 110 | 127 | 144 | 160 |
| (e) | 282 | 323 | 363 | 401 |
| (f) | 221 | 237 | 252 | 266 |

Table 5 Gibbs free energy change $\Delta G_T$ of the reactions in Table 1 calculated using standard formation molar Gibbs free energy method, the unit is kJ mol$^{-1}$

| ordinal | $\Delta G_{700}$ | $\Delta G_{800}$ | $\Delta G_{900}$ | $\Delta G_{1000}$ |
|---|---|---|---|---|
| (a) | 266 | 265 | 264 | 263 |
| (b) | 145 | 145 | 145 | 145 |
| (c) | -5.55 | 10.7 | 26.8 | 42.7 |
| (d) | 116 | 131 | 146 | 161 |
| (e) | 94.4 | 104 | 113 | 122 |
| (f) | 216 | 225 | 233 | 241 |

The $\Delta G_T$ data listed in Tables 4 and 5 were illustrated in Fig. 1 to conduct a comparison analysis. It could be observed that the $\Delta G_T$ values calculated using two methods were generally similar, especially for reaction (a). As mentioned in section 1.2, there was a certain deviation of the calculated results from the real values using the Gibbs free energy function method for reactions (e) and (f), which involved $Ba(OH)_2$. However, the results of reactions (e) and (f) using two methods were similar, as shown in Fig. 1.

Besides, $\Delta G_T$ values were positive except reaction (c), indicating that most reactions could hardly occur under the standard condition. For reaction (c), $\Delta G_T$ increased along with $T$ raising. $\Delta G_T$ was negative at 700 K, and positive at higher $T$ of 900 K and 1000 K, according to both calculation methods.

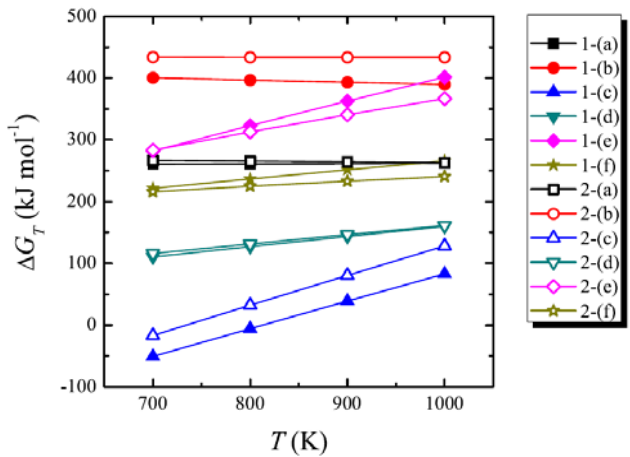

Fig.1 Gibbs free energy change $\Delta G_T$ values of the reactions (a)-(f) in Table 1, calculated using Gibbs free energy function method and standard formation molar Gibbs free energy method, which were labeled as 1 and 2 respectively.

**2.2 Conversion of BaF₂ and BaCO₃**

The negative $\Delta G_T$ at 700 K (approximately 427°C) could account for the existence of $BaCO_3$ phase in the samples quenched at low temperatures, which were prepared using precursors with very low fluorine amount. In our previous study [6] using precursor with 10.3% fluorine amount relative to conventional TFA-MOD process, $BaCO_3$ phase could be detected by the attenuated total reflectance Fourier transform infrared spectra at 400°C. In the study of Jin et al. [12], using precursor with 7.7% fluorine amount, $BaCO_3$ phase could be detected by X-ray diffraction at 400°C and 500°C.

Moreover, in the above two studies [6, 12], $BaCO_3$ was not detected at higher temperatures. Such a behavior might be attributed to two reasons. First, $\Delta G_T$ of reaction (c) increased along with $T$ raising and became positive at higher $T$, leading to the conversion of $BaCO_3$ to $BaF_2$ through reaction (c) in the backward direction. Second, there was an oxygen flow in the furnace in the MOD heat treatment which could remove the $CO_2$ out of the film continually, also leading to the conversion reaction of $BaCO_3$ to $BaF_2$.

## 3 Summary

In this paper, two methods, Gibbs free energy function method and standard formation molar Gibbs free energy method, were used to calculate the Gibbs free energy change ($\Delta G_T$) at temperatures ranging from 400 K to 1000 K, to analyze the possible conversion reactions of $BaF_2$. $\Delta G_T$ values calculated using the two methods were generally similar. $\Delta G_T$ of the reaction $BaF_2$ converting to $BaCO_3$ was found to be negative at 700 K and positive at high temperatures, consistent with the studies of very low fluorine amount precursors. This study could verify the $BaF_2$ phase stability at higher temperatures and could help to investigate the evolution of intermediate phase in heat treatment of MOD-YBCO process.

### References

[1] Holesinger T. G, Civale L, Maiorov B, et al. Progress in Nanoengineered Microstructures for Tunable High-Current, High-Temperature Superconducting Wires. Advanced Materials, 2008, 20(3): 391-407.

[2] Hu R, Dong H, Li J, et al. Study on Microstructure Characterization of YBCO Bulk Prepared by Directional Seeded Infiltration-Growth. Rare Metal Materials & Engineering, 2008, 37(5): 854-858.

[3] Jin L H, Lu Y F, Feng J Q, et al. Development of modified TFA-MOD approach for $GdBa_2Cu_3O_y$, film growth. Materials Letters, 2013, 94(3): 23-26.

[4] Zhao X, Gao C, Xia Y, et al. Preparation of YBCO Coated Conductors on RABiTS Substrate with Advanced TFA-MOD Method. Rare Metal Materials & Engineering, 2011, 40(s3): 342-345.

[5] Huang R, Feng F, Wu W, et al. A water-free metal organic deposition method for $Yba_2Cu_3O_{7–\delta}$ thin film fabrication. Superconductor Science & Technology, 2013, 26(11): 115010.

[6] Wu W, Feng F, Zhao Y, et al. A low-fluorine solution with a 2:1 F/Ba mole ratio for the fabrication of YBCO films. Superconductor Science & Technology, 2014, 27(5): 105-112.

[7] Li M, Yang W, Shu G, et al. Controlled-Growth of $Yba_2Cu_3O_{7–\delta}$ Film Using Modified Low-Fluorine Chemical Solution Deposition. IEEE Transactions on Applied Superconductivity,

2015, 25(3): 1-4.

[8] Goyal A, Second-generation HTS conductors, Boston: Kluwer Academic Publishers, 2005: Chapter 11, 179-194.

[9] Ye D L et al. Thermodynamics data handbook of practical inorganic materials, 2nd Edition, Beijing: Metallurgical Industry Press, 2002: 1-1209 (in Chinese).

[10] Klotz I M, Rosenberg R M, Chemical thermodynamic: basic concepts and methods, 7th Edition, Wiley: Hoboken, N.J., 2008: 1-110.

[11] Barin I, Thermochemical data of pure substances, 3rd Edition, Weinheim: WILEY-VCH Verlag GmnH, 1995: 1-1885.

[12] Jin L H, Li C S, Feng J Q, et al. Optimization of fluorine content in TFA-MOD precursor solutions for YBCO film growth. Superconductor Science & Technology, 2016, 29(1): 015001.